\documentclass{JAC2003}
\usepackage{amsmath}
\usepackage[utf8]{inputenc}
\usepackage{graphicx}
\usepackage{psfrag}
\usepackage{times}
\usepackage[T1]{fontenc}
\usepackage{xspace}

\def\j{\ensuremath{\text{j}}} 

\def\vecfunc#1{\ensuremath{{#1}}}

\def\FEST3D{FEST~3D\xspace}

\usepackage{alphalph}
\makeatletter
\newcommand*{\fnsymbolsingle}[1]{%
  \ensuremath{%
    \ifcase#1%
    \or *%
    \or \ddagger
    \or \mathsection
    \or \mathparagraph
    \or \dagger
    \else
      \@ctrerr
    \fi
  }%
}
\makeatother
\newalphalph{\fnsymbolmult}[mult]{\fnsymbolsingle}{}

\title{MODELING MICROWAVE/ELECTRON-CLOUD INTERACTION} 
\author{M. Mattes\thanks{michael.mattes@epfl.ch}, E. Sorolla\thanks{eden.sorolla@epfl.ch}, EPFL-STI-IEL-LEMA, Station 11, 1015 Lausanne, Switzerland\\
F. Zimmermann\thanks{frank.zimmermann@cern.ch}, CERN, Accelerators and Beam Physics Group, 1211 Geneva 23, Switzerland}

\date{}
\begin{document}

\maketitle
%
\begin{abstract}
Starting from the separate codes BI-RME and ECLOUD or PyECLOUD,
we are developing a novel joint simulation tool,  
which models the combined effect of a charged  
particle beam and of microwaves on an electron cloud. 
Possible applications include   
the degradation of microwave transmission in tele-communication satellites by electron clouds;  
the microwave-transmission tecchniques  
being used in particle accelerators for the purpose of electron-cloud diagnostics;  
the microwave emission by the electron cloud itself in the presence of a magnetic field; and 
the possible suppression of electron-cloud formation in an accelerator by injecting microwaves  
of suitable amplitude and frequency. 
A few early simulation results are presented.  
\end{abstract}

\section{Motivation}  
Electron multiplication on surfaces exposed to either an oscillating electromagnetic field or to a 
pulsed electric field gives rise to the phenomenon of multipacting, which can significantly 
degrade the performance of radiofrequency devices operating in high-vacuum conditions. 
These include, for example, components for satellite tele-communication 
like waveguide filters but also accelerating cavities for particle accelerators.  
In both examples, the electron multiplication can result in a quasi-stationary ``electron cloud'' 
inside the devices perturbing their performance.

Though studies of electron multipacting in microwave devices as well as of 
beam-induced multipactor in particle accelerators have both been ongoing separately 
for several decades already \cite{ecloud1,ecloud2,ecloud3,microwave}, 
little is known about the simultaneous interaction of an electron 
cloud with both microwaves and a particle beam \cite{ecloud02,caspers}. 
This lack of knowledge has so far prevented the proper interpretation of microwave 
transmission measurements which have been applied in several accelerators afflicted by an electron cloud. 
A related phenomenon is encountered in satellite-based telecommunication systems, which nowadays suffer from the 
problem of an electron-cloud build up interacting with the microwave field used to transmit information.

\section{Model}  
In the associated oral contribution at ECLOUD12 we have outlined the model applied 
to describe the interaction between electrons and microwaves. 
Differently from \cite{bib4}, the modelling is based on time-domain Green functions describing the radiation of a single electron represented 
as a point charge.  Using the modal expansion representation of the involved Green functions, we are able to take into account 
the exact boundary conditions of the devices, e.g.~waveguides or cavities.

\section{Example Results}  
\begin{figure}
 \centering
 \includegraphics[width=0.4\textwidth]{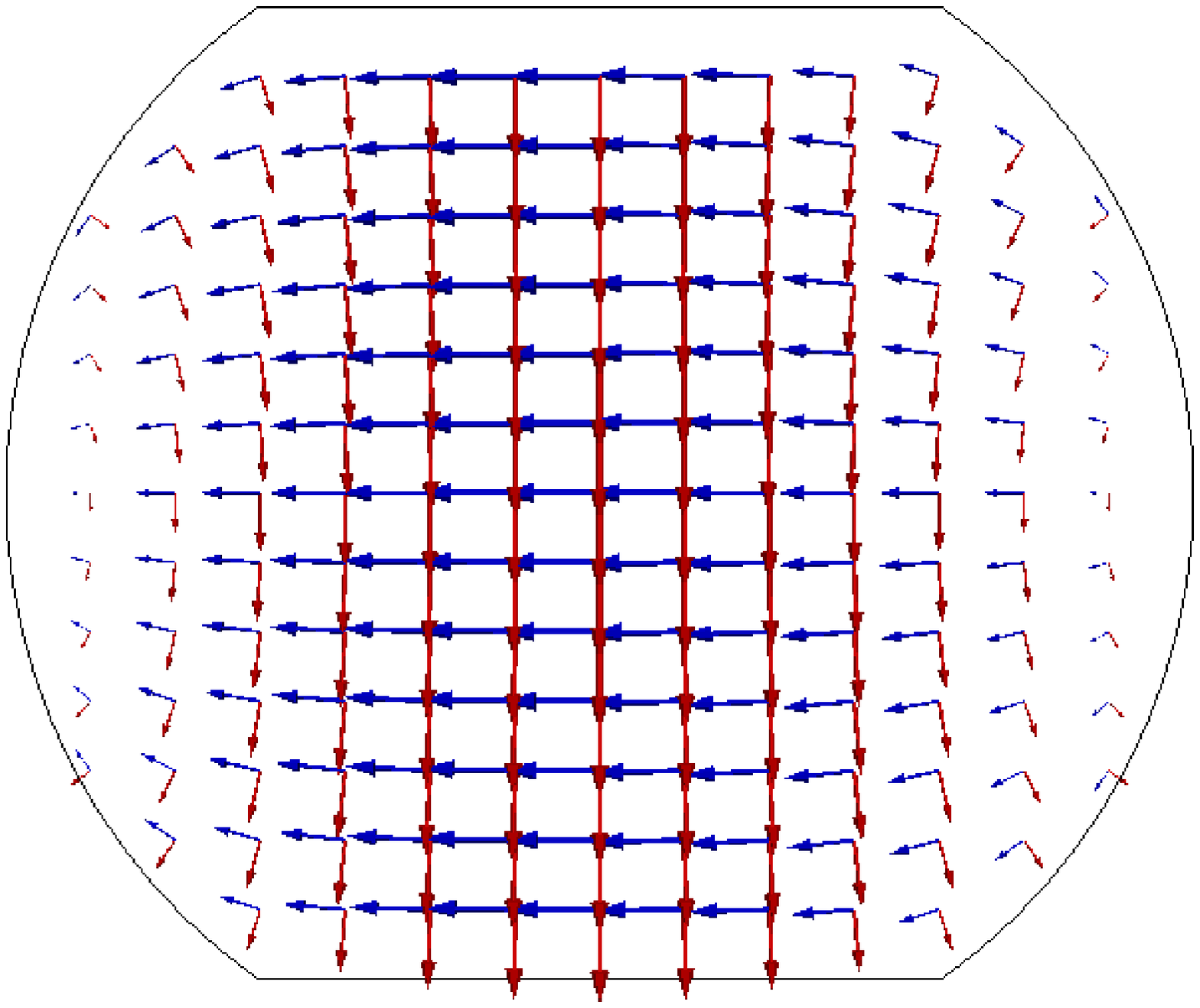}
 \includegraphics[width=0.4\textwidth]{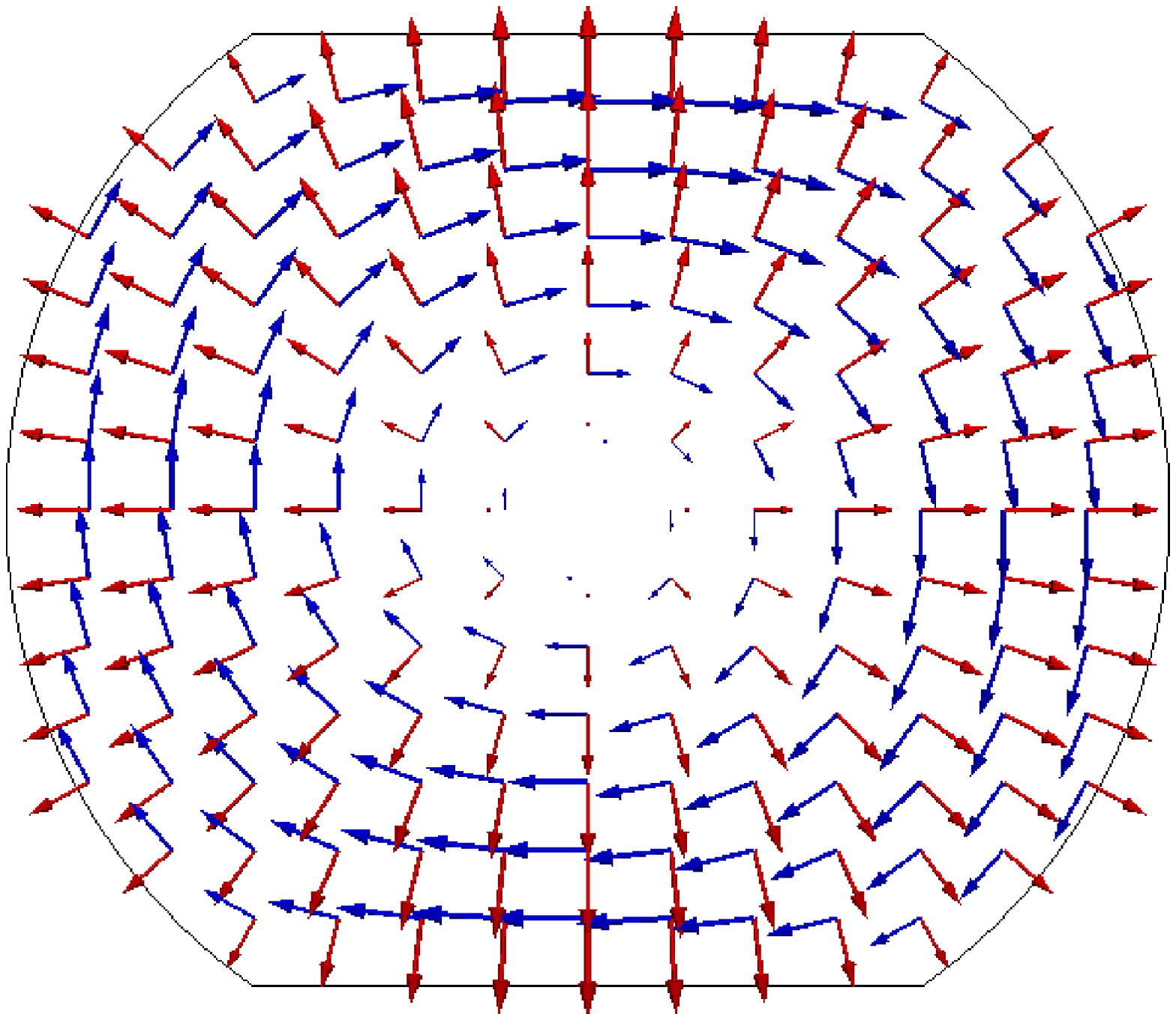}
 \caption{\label{fig:fig1}First TE (top) and TM (bottom) mode of LHC beam pipe. 
The red arrows represent the transverse electric field, the blue ones the transverse magnetic field.}
\end{figure}

Figure~\ref{fig:fig1} shows the example of the LHC beam pipe representing the transverse electric (red arrows) 
and magnetic field (blue arrows) of the first TE and TM mode, respectively. The modes are computed using the BI-RME algorithm \cite{bib5}.

\begin{figure}
\centering
\psfrag{t [ns]}[t][][0.6]{$t$ [ns]}
\psfrag{z [mm]}[t][][0.6]{$z$ [mm]}

\psfrag{FFT(dA/dt) [V/m/Hz]}[][][0.6]{$\j\omega \mathcal{F}\{\vecfunc A\}$ [V/Hz]}
\psfrag{f [GHz]}[t][][0.6]{$f$ [GHz]}

\includegraphics[width=0.45\textwidth]{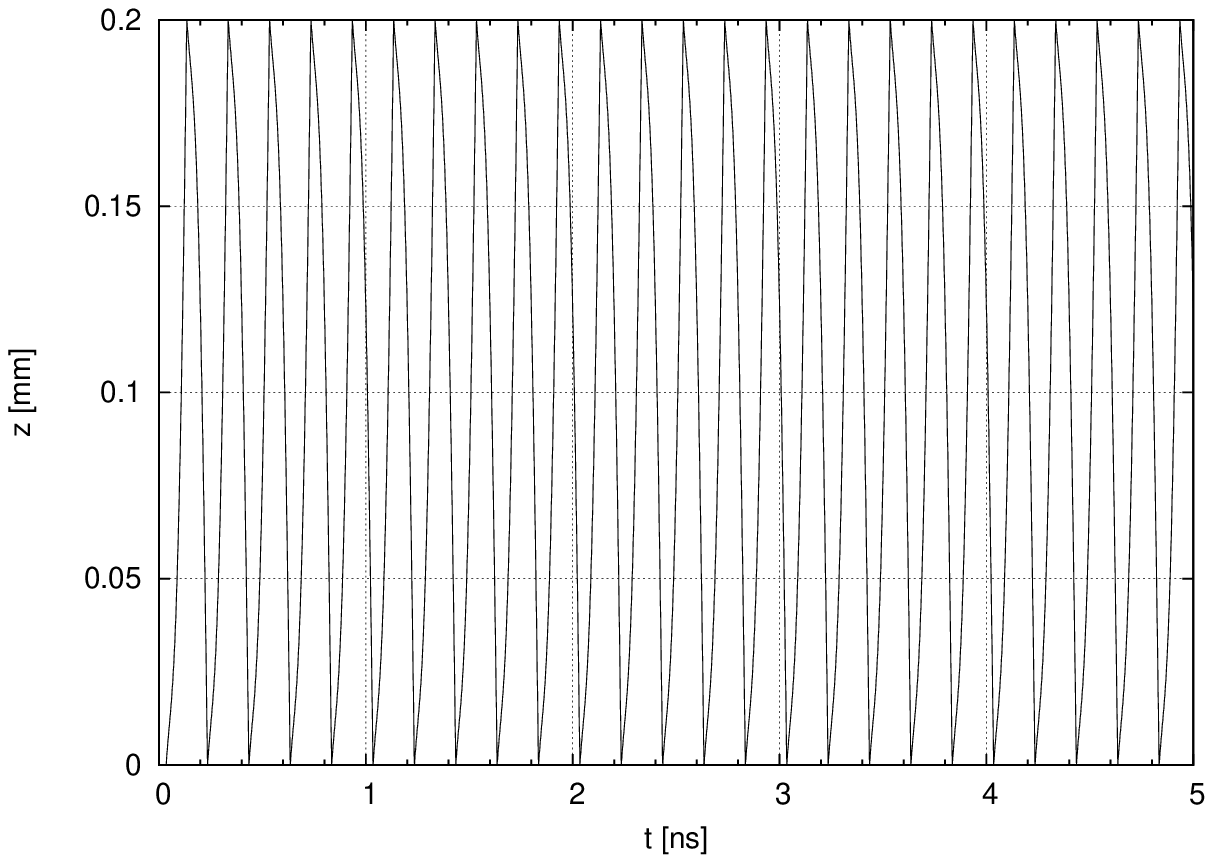}

\includegraphics[width=0.45\textwidth]{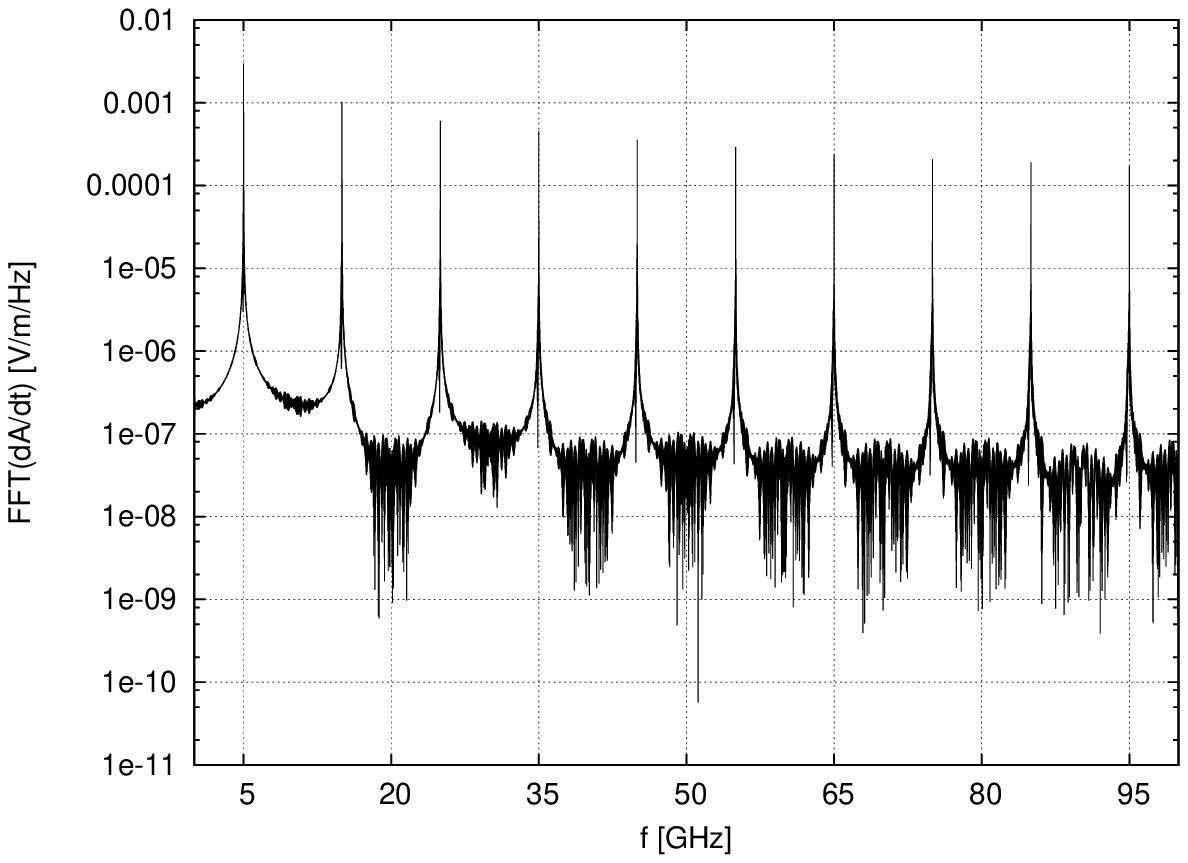}
\caption{\label{fig:fig2}Multipacting electron (top) and the associated radiated electric field spectrum (bottom) 
inside a parallel-plate waveguide of height 0.2mm. Simulation parameters: RF frequency: 5\,GHz; RF voltage: 70\,V, 
initial velocity of electrons: 2\,eV.}
\end{figure}

In Fig.~\ref{fig:fig2} the example of a single electron under multipactor regime between two parallel plates separated by 0.2\,mm is depicted. 
The electric field, driving the electron, oscillates at 5\,GHz. The upper figure shows the position of the electron, the lower one 
the part of the radiated electric field density related to the time-derivative of the vector potential $\vecfunc A$. The spectrum 
includes all harmonics that are odd multiples of the fundamental one and are generated by the moving electron.

\begin{figure} 
\centering
\psfrag{t [us]}[t][][0.6]{$t$ [$\mu\text{s}$]}
\psfrag{sigma}[t][][0.6]{{electron density [$\mu\text{C}$/$\text{m}^2$]}}
\psfrag{dA/dt}[t][][0.6]{{$\partial \vecfunc A/\partial t$ [V]}}
\psfrag{f [GHz]}[t][][0.6]{$f$ [GHz]}
\psfrag{FFT(dA/dt) [V/m/Hz]}[][][0.6]{$\j\omega \mathcal{F}\{\vecfunc A_1\}$ [V/Hz]}
\psfrag{Pmean [W/GHz]}[][][0.6]{$P_{\text{rms}}$ [W/GHz]}

\includegraphics[width=0.45\textwidth]{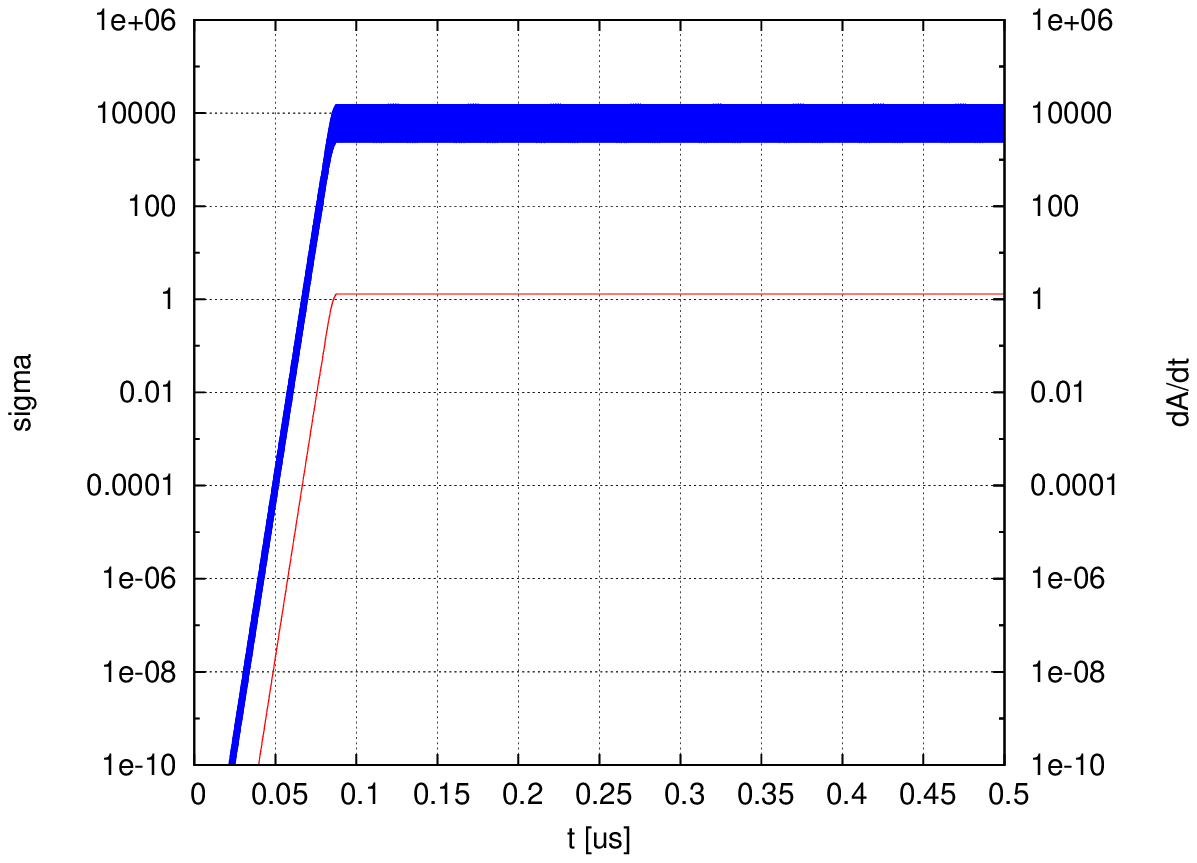}

\psfrag{sigma}[t][][0.8]{{space charge [nC]}}
\includegraphics[width=0.45\textwidth]{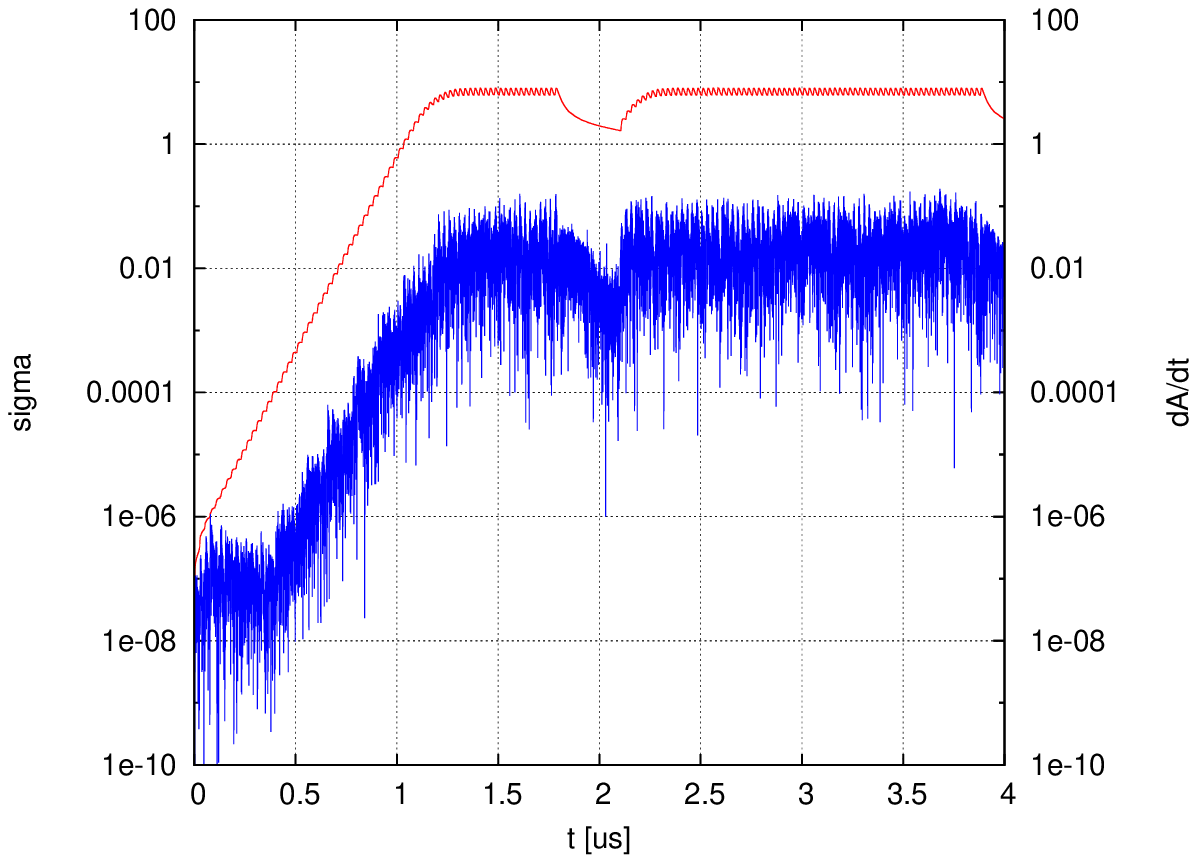}

\caption{\label{fig:fig3}Example of an electron cloud build-up inside the parallel plate waveguide of Fig.~\ref{fig:fig2} 
(top) and in the LHC beam pipe. The red curve repreesents the electron density (top) respectively the space charge (bottom). 
The blue lines are the associated time derivatives of the vector potential$\vecfunc A$, thus the electric field.}
\end{figure}

The model developed is useful for microwave diagnosis of devices prone to microwave breakdown due to multipaction. 
The upper picture of Fig.~\ref{fig:fig3} shows the evolution of the electron density for the case of the parallel 
plate waveguide of Fig.~\ref{fig:fig2} applying the model of \cite{bib6}. 
The blue curve represents the electron density, the red one the associated radiated electric field contribution from 
the time-derivative of the vector potential $\vecfunc A$. There is a clear correlation between the two curves. 
The lower part of Fig.~\ref{fig:fig3} is an example from the LHC beam pipe. The space charge build up was simulated with 
PyECLOUD~\cite{pyecloud} (blue curve). Based on the electron trajectories the radiated electromagnetic field has been computed (red curve). 
Again, a correlation of the two curves can be observed.

\section{ACKNOWLEDGMENT}
This project has been financed by the Swiss National Science Foundation (SNF) under Grant 
No.~200021\_129661 ``Modelling microwave-electron interaction in the LHC beam pipe''.

\end{document}